# Optimization of deep learning models for the prediction of gene mutations using unsupervised clustering

**Author:** Zihan Chen[1,&], Xingyu Li[1,&], Miaomiao Yang[2], Hong Zhang[1,*], Xu Steven Xu[3,*]

[1] Department of Statistics and Finance, School of Management, University of Science and Technology of China;
[2] Clinical Pathology Center, The Fourth Affiliated Hospital of Anhui Medical University, Hefei, Anhui, China;
[3] Data Science/Translational Research, Genmab Inc., Princeton, New Jersey, USA;

**Corresponding author: Hong Zhang**, Department of Statistics and Finance, School of Management, University of Science and Technology of China, Hefei, Anhui 230026, China

E-mail : zhangh@ustc.edu.cn

**Steven Xu,** Data Science/Translational Research, Genmab Inc., Princeton, New Jersey, USA

E-mail: sxu@genmab.com

---

[1] &These authors contributed equally to this work and should be considered co-first authors

**Abstract**

Deep learning models are increasingly being used to interpret whole-slide digital pathology images (WSIs) and to predict genetic mutations. Currently, it is commonly assumed that tumor regions have most of the predictive power. However, it is reasonable to assume that other tissues from tumor microenviroment may also provide important predictive information. In this paper, we proposed an unsupervised clustering-based multiple-instance deep learning model for the prediction of genetic mutations using WSIs of three cancer types obtained from The Cancer Genome Atlas. Our proposed model facilitated the identification of spatial regions related to specific gene mutations and exclusion of patches that lacked predictive information through the use of unsupervised clustering. This resulted in a more accurate prediction of gene mutations when compared with models using all image patches on WSIs and two recently published algorithms for all the three different cancer types evaluated in this study. In addition, our study validated the hypothesis that the prediction of gene mutations solely based on tumor regions on the WSI slides may not always provide the best performance. Other tissue types in the tumor micro-environment could provide a better prediction ability than tumor tissues alone. These results highlight the heterogeneity in the tumor microenvironment and the important of identification of predictive image patches in digital pathology prediction tasks.

# 1. Introduction

The diagnosis of cancer is typically based on a histopathological assessment of tissue sections, and supplemented by genetic and other molecular tests (Abeshouse et al., 2015; Bailey et al., 2016; Dienstmann et al., 2017; Lindeman et al., 2013; Russnes et al., 2017; Woodman et al., 2012). 错误!未找到引用源。错误!未找到引用源。错误!未找到引用源。错误!未找到引用源。 The identification of molecular biomarkers and gene mutations is becoming increasingly important for the development of novel treatment options. For example, the KRAS mutations, present in about 30% to 50% of colorectal cancers (CRC), are associated with poor prognosis and advanced disease (Bazan et al., 2002; Castagnola and Giaretti, 2005; Liu et al., 2011; Nosho et al., 2008; Poehlmann et al., 2007; Russo et al., 2005; Suehiro et al., 2008). In lung adenocarcinoma (LUAD), EGFR has been reported to be mutated in about 20% of patients. As a result, multiple EGFR therapies aimed at targeting these mutations have been developed and approved by the Food and Drug Administration (Blumenthal et al., 2017; Pérez-Soler et al., 2004). However, due to the long turnaround time, tissue usage and costs in the current oncology workflows for genetic mutations from tissue samples (Rusch et al., 2018), there is a growing need for the development of cheap, scalable fast alternatives to predict genetic mutations.

Various deep learning-based computer vision algorithms have been developed to predict gene mutations whole-slide images (WSIs) (Chen et al., 2020; Ding et al., 2020; Kather et al., 2020; Liao et al., 2020; Qu et al., 2021; Srinidhi et al., 2021; Wang et al., 2021). Coudray et al. (2018) proposed a deep convolutional neural network (DeepPATH) to predict gene mutations in LUAD based on WSIs. 错误!未找到引用源。 Kather et al. (2020) proposed, optimized, and extensively validated a one-stop-shop workflow based on the lightweight neural network, ShuffleNet. They showed that a wide range of genetic mutations, molecular tumor subtypes, gene expression signatures, and standard pathology biomarkers could be inferred from WSIs.

Since a large number of image patches (ranging from hundreds to thousands) are available for each WSI, and not all areas within the WSIs are relevant to gene mutations, direct use of all image patches of a WSI prediction model may limit the prediction performance of the model. It has therefore been postulated that certain image regions or patches within the WSI (e.g., tumor) could carry more predictive value. To overcome this issue, pathologists annotated tumor regions relevant to the diagnostic task have been used to train predictive models (Cheng et al., 2018; Gurcan et al., 2009; Kather et al., 2020; Levy et al., 2020; Sirinukunwattana et al., 2021; Yuan et al., 2012; Zhu et al., 2016a; Zhu et al., 2016b). Scientists also trained tissue classifiers (tumor and nontumor) to automatically select tumor-like tiles to predict mutated genes in different cancers (Bilal et al., 2021; Coudray et al., 2018; Jang et al., 2021; Qu et al., 2021).

In the field of digital pathology, unsupervised clustering has been widely used to reduce the dimensionality of patches to facilitate multiple instance learning (e.g., patches from WSIs can be fit on a GPU at once) (Abbet et al., 2020). This method was also used to derive additional cluster-based features, and to identify rare events. Dooley et al. (2018) and Zhu et al. (2019) clustered patches into different clusters and used the frequency of patches in each cluster as a new feature to predict heart transplant rejection. Similarly, 错误!未找到引用源。 Abbet (Abbet et al., 2020) proposed a self-supervised learning method that jointly learns from a representation of tissue regions as well as a clustering metric to identify spatial tissue features such as cluster probabilities and cluster transition probabilities.

In addition, unsupervised clustering has been commonly used in image-based deep-learning survival analysis. Yao et al. (2020) clustered the patches in each WSI individually into different phenotype clusters. One patch from each cluster was then sampled and was used to predict survival in CRC patients. 错误!未找到引用源。 Sharma et al. (2021) deployed a local cluster-based (clustering patches from a single WSI) sampling approach for identifying children with celiac disease. Zhu et al. and Yue et al. used global clustering of patches from all patients to train a survival model based on the information derived for each cluster. The features from the most predictive clusters were then aggregated across the patches from each cluster to predict the outcome. Muhammad et al. (2021) used patch features grouped by global centroids to calculate the local slide-level centroid and concatenated the nearest patches to local centroids to represent each slide. The model was then trained with survival data. Their approach performed better than other approaches used in the modeling of intrahepatic cholangiocarcinoma.

Although various applications have been developed for unsupervised clustering in digital pathology, very few studies evaluated the use of unsupervised clustering for the identification of image patches linked with genetic mutations. Therefore in this paper, we proposed an unsupervised clustering-based multiple-instance learning method to develop a deep-learning model for optimization of the prediction of genetic mutations using the WSIs of three common cancer types obtained from The Cancer Genome Atlas (TCGA).

## 2. Methods and materials

### 2.1. TCGA dataset

Datasets of WSIs for three tumor types, including colorectal (CRC), head and neck squamous cell carcinoma (HNSCC), and lung adenocarcinoma (LUAD), were retrieved from TCGA available on https://portal.gdc.cancer.gov. The corresponding TCGA gene mutation data and subtype data were downloaded from the [https://xenabrowser.net/datapages/](https://xenabrowser.net/datapages/) website.

### 2.2. Clinically relevant gene mutations

Clinically relevant genes for each cancer type reported in Refs (Mosele et al., 2020; Network, 2015) were selected for analysis (Table 1). We assigned each patch with label 1 or 0, depending on the presence or absence of the mutation in that patient.

### 2.3. Image preprocessing

The background region with no tissue from each hematoxylin and eosin (H&E) stained WSI was excluded using an adapted Otsu method (1979). This technique involves separating the pixels in each image on the greyscale space into foreground and background. The background was then removed leaving only the tissue. The tissue areas of the image were then tiled into small non-overlapping patches, each with a dimension of 224 × 224 pixels. Macenko's method (2009) was then used to normalize the color patches synchronously (Supplementary Figure 1). For the LUAD, HNSCC, and CRC, the number of patches extracted from the WSIs ranged from about 100 to 50,000 (average = 12,664), 100 to 30,000 (average = 12,772), and a few hundreds to 30,000 (average = 7,888), respectively.

## 2.4. Feature extraction

We employed a fine-tuned Xception model to extract features from the image patches (Li et al., 2022), which was pre-trained on the ImageNet datasets and fine-tuned on the CRC dataset (Kather et al., 2019). A feature vector with a dimension of 256 was extracted from each patch. For each patient, an n × 256 feature matrix was obtained whereby n represents the number of patches in the patient of interest.

## 2.5. Unsupervised clustering

We randomly selected 100 patients from each of the three tumor types (LUAD, HNSCC, and CRC) in the TCGA dataset. After pooling all patches from the 100 patients of each cancer type, we used K-means clustering to cluster these patches into four groups. A k-NN algorithm was used to assign cluster labels to the rest of the patches of that cancer type, which were not included in the process of building the clustering model.

To our knowledge, this is the first study that has made use of unsupervised clustering to optimize the prediction of genetic mutations on WSIs. However, we leveraged studies currently available in other areas to select the number of clusters (Sharma et al., 2021). Based on the trade-off between computational efficiency and clustering performance on the deep learning platform, we decided to use a cluster number of four (k=4).

## 2.6. Best-cluster optimized multiple-instance learning

To study the effect of clustering, on each cluster we trained a patch-level multilayer perceptron (MLP) classifier that used the features of the patches only in each cluster as the input to estimate the mutation probability of each patch. The Adam algorithm was used to optimize the cross-entropy. After averaging the predicted probability, we obtained a classifier for each slide level. The algorithm was then tested on WSIs obtained from the TCGA dataset. Figure 1 shows the pipeline method used to develop our model.

**2.7. Performance evaluation of the best-cluster optimized method**

The performance of the best-cluster optimized method for the prediction of genetic mutations was compared with (1) a WSI-based approach without unsupervised clustering, (2) a tumor region based method (CRC only), and (3) other published algorithms as follows.

**2.7.1. Comparison to the WSI-based approach without unsupervised clustering**

As a benchmark comparison, we also trained the MIL classifier using all patches from patients as input without clustering the patches. All the patches obtained from the WSIs were used to train a patch-level network. The average predicted probability of patches was used to predict the slide-level mutation (Coudray et al., 2018).

**2.7.2. Comparison to the prediction model based on tumor regions**

Patches from tumor areas have often been used to train prediction models for gene mutations (Bilal et al., 2021; Coudray et al., 2018; Jang et al., 2021; Qu et al., 2021). For CRC, we selected tumor tiles using a fine-tuned Xception-based tissue-type classifier (Li et al., 2022). We then trained the mutation prediction model on the tumor patches and compared the performance of the model with the best-cluster-based model. Since patch-level labels for tissue types were only available for the CRC dataset, we evaluated the performance of the two methods only on the CRC dataset.

### 2.7.3. Comparison to other published baseline algorithms

In addition, we compared our method with two recently published baseline methods by Dooley et al. (2018) and Zhu et al. (2017) that also utilized unsupervised clustering to improve the accuracy of their prediction models. Dooley et al. (2018) used unsupervised clustering to group patches into different clusters and then used the frequency of patches of each cluster as a new feature for classification. On the other hand, the algorithm developed by Zhu et al. (2017) used unsupervised clustering to facilitate the development of a deep-learning survival model. This method makes use of a K-means cluster instead of manually annotated regions to select the effective cluster for the final prediction.

The method proposed by Zhu et al. was adapted to develop our algorithm as follows. We first performed global clustering of the patches from all patients. We then trained a separate MLP classifier for each cluster. Finally, the average area under the curve (AUC) for predicting the gene mutation was obtained using a 5-fold cross validation technique. Clusters with an AUC greater than 0.5 were selected and the weights for cluster $j$ in patient $i$ were calculated as follows,

$$w_{ij} = \frac{n_{ij}}{n_i}, i \in \{1, \dots, N\}, j \in \{1, \dots, J\},$$

whereby $n_{ij}$ represents the number of patches belonging to cluster j in patient $i$ and $n_i$ represents the number of total patches extracted from patient $i$

Similar to the Zhu et al. method, the features for that patient in the selected cluster were calculated as,

$$x_{ij} = w_{ij} \sum_{k=1}^{k=K} x_{ijk}/K,$$

where $x_{ij}$ represents the output features in cluster $j$ for patient $i$. We then concatenated the weighted feature from the selected clusters and used it to train a support vector machine (SVM) classifier for genetic mutation prediction.

3. Experiments

In all experiments, we used stratified five-fold cross validation according to the mutation status for each gene whereby the dataset obtained from the TCGA for each cancer type was divided into five folds to avoid the data imbalance problem. In each fold, 80% of the data were used for model training and 20% of the data were used to test the performance of the model. During training, we used the Adam optimization method with an initial learning rate of 0.00005 and a cosine annealing schedule with a maximum number of 20 iterations. Training and validation were done over 1,000 iterations. The performance of the model was evaluated by calculating the AUC of a receiver operator characteristic (ROC) curve. When calculating the cross-entropy loss, we assigned more weight to classes with a small number of training images so that the network was punished more if it falsely predicted the labels of these classes.

4. Annotation of image patches

Automatic and semi-automatic methods were used to classify the WSIs obtained from the TCGA database.

**4.1. Automated annotation with tissue-type classifier for the CRC WSIs**

Kather et al. developed a deep-learning classifier to classify CRC image tiles into eight tissue types: adipose tissue (ADI), background (BACK), debris (DEB), lymphocytes (LYM), mucus (MUC), smooth muscle (MUS), normal colon mucosa (NORM), cancer-associated stroma (STR), and colorectal adenocarcinoma epithelium (TUM) (Li et al., 2021). The pathologist annotated NCT-CRC-HE-100K data provided by Kather et al. (Li et al., 2022) was used to train a similar tissue-type classifier model while the CRC-VAL-HE-7K image sets were used to validate the model. The overall accuracy of the tissue-type classification model was 99% for the training dataset NCT-CRC-HE-100K and 94.4% for the validation image set CRC-VAL-HE-7K (Supplementary Figure 2). The tissue type of each image tile from the TCGA dataset was predicted using the fine-tuned Xception based tissue type classifier.

**4.2. Semi-automatic annotation**

For the LUAD, HNSCC, as well as CRC cohorts, we developed semi-automatic procedures to facilitate the annotation of the image patches from each cluster. For each cluster of the four clusters within the LUAD, HNSCC, and CRC cohorts, additional K-means clustering (k =4) was performed. Then, we selected four neighboring patches around the center of each of the four subclusters for the pathologist to annotate. In total, the pathologist annotated 64 patches for each of the three cancer types out of a total of 5,496,176 patches for LUAD, 5,504,732 patches for HNSCC, and 3,265,632 patches for CRC. We used the CRC data to confirm that the semi-automatic annotation approach could identify similar tissue types to those identified by the tissue-type classifier for CRC.

## 5. Results

### 5.1. Composition of the tissue clusters within the LUAD, HNSCC, and CRC datasets

Figure 3 shows that K-means clustered the tiles into four distinct clusters for the three TCGA datasets (LUAD, HNSCC, and CRC). Figures 4-6 show the image tiles representing the most common tissue type among the four neighborhood patches near the center of each subcluster (four subclusters for each cluster) for the LUAD, HNSCC, and CRC, respectively. For the LUAD, cluster 2 mainly consisted of tumor tissues, while cluster 4 primarily included stromal cells. Clusters 1 and 3 consisted of a mix of red blood cells, stromal cells, pulmonary alveolus, tumor, lymphocytes, proliferating fibroblasts, and other non-tumoral cells (Figure 4). For the HNSCC cohort, clusters 3 and 4 consisted of the non-tumor and tumor compartments, respectively. Cluster 1 from the HNSCC cohort was a mix of lymphocytes and tumor cells, while cluster 2 comprised mostly non-tumor cells with some tumor cells (Figure 5).

For CRC, we examined the tissue types of each cluster using the CRC tissue type classifier (Li et al., 2021). Figures 6 and 7 show that cluster 1 consisted mainly of tumor and mucin cells, while almost all patches in Cluster 2 were tumor cells. Cluster 3 of the CRC primarily included muscular and stromal cells with some debris and tumor cells as well. Various tissue types were present in cluster 4, which included all 8 tissue types (plus a small number of background patches) in the classifier (Figure 7). It is worth mentioning that the annotations based on the tissue classifier were generally consistent with the manual annotations provided by the pathologist (Figures 6a and 6b).

## 5.2. Prediction of gene mutations by tissue clusters

Tables 2-4 illustrate the average AUC values obtained from the five-fold cross validation using the three TCGA datasets (LUAD, HNSCC, and CRC, respectively) for the four prediction models based on the image tiles from the four individual clusters. The image tiles from the different clusters had different predictive abilities. For the LUAD (Table 2), the tumor cells in cluster 2 provided the best prediction for the TP53 and STK11 mutations, suggesting that the mutant-like image features for TP53 and STK11 are mainly found within the tumor region (refer to the heatmap in Figure 8). This finding is consistent with the results obtained by Coudray et al. (2018). The tumor patches also predicted the EGFR mutations well (Table 2). The stromal cells in cluster 4 provided the highest AUC for the prediction of the ALK gene mutation (Table 2) and the image tile with the highest likelihood of ALK mutation demonstrated stromal features (Figure 8). Models based on image tiles from clusters 1 and 3 which consisted of a mix of red blood cells, stromal cells, interalveolar septum cells, and other non-tumoral cells provided the best prediction for the KEAP1 and KRAS mutations, respectively.

Similarly, in the HNSCC cohort, a remarkable difference in the AUC was noted for the prediction of gene mutations for the different clusters (Table 3). The predictive performance of the tumor cells in cluster 4 was very high for the TP53, HRAS, CASP8, and NSD1 mutations. The heatmap in Figure 9 shows that the TP53 mutant-like features are highly present in the tumor compartment of HNSCC. Conversely, non-tumor cells in cluster 3 provided the best prediction performance for the NSD1 mutations, while the mix of lymphocytes and tumor cells in cluster 1 was better at predicting the CASP8 mutation. The image tiles with the highest likelihood of predicting the NSD1 and CASP8 mutations were non-tumoral cells (Figure 9). The models based on the non-tumor cells (including blood vessels, debris, etc.) in cluster 2 outperformed the other three clusters in terms of the prediction for the DNAH5, HRAS, and PTEN mutations. Consistently, the image tiles with the highest likelihood of DNAH5, HRAS, and PTEN consisted of red blood cells, stroma/red blood cells, and tumor/blood cells, respectively (Figure 9).

For CRC (Table 3), features from both cluster 1 (primarily tumor and mucin tissues) and cluster 4 (a mixture of all types of tissues) had the best predictive performance for the vast majority of the genes, although the tumor compartment (cluster 2) also had a relatively good prediction performance. These results suggest that the mutant-like image features for these clinically relevant genes in CRC are not exclusively confined to the tumor regions (Figure 10). Other tissues, particularly mucin, also have a great predictive value for predicting genetic mutations on CRC WSIs. These findings are consistent with the work of Nguyen et al., whereby image patches for tumor and mucus regions tended to better predict the MSI status for CRC patients (Nguyen et al., 2022). In addition, the image tiles in cluster 4 consisting of lymphocytes had the highest predictive ability for the ERBB2, ATM, and MET mutations, while an image tile with normal-tissue-like features had the highest likelihood of predicting the PIK3CA mutation (Figure 10).

**5.3. Mutation prediction comparison between the best-cluster optimized model and the**

**WSIs method**

We used the average AUC from the five-fold cross-validation to compare our best-cluster-based approach with the model using all patches from a WSI, and we focused on genes with an average AUC greater than 0.6 (Figure 11). The cross validation showed that the selected best cluster consistently provided an improvement in the prediction of genetic mutations for all the three cancer types (i.e., LUAD, HNSCC, and CRC) (Figures 11). The AUC of our proposed model was on average 0.08 higher when compared with the WSI method, which indicates an overall higher prediction performance (Figure 11b). In the LUAD cohort, a remarkable improvement in the AUC (ΔAUC) was observed for the STK11 (0.061), ALK (0.051), and KRAS (0.046) mutations. For the HNSCC cohort, an improvement in the AUC was noted for the DNAH5 (0.083), HRAS (0.067), and PTEN (0.057) mutations, while the prediction of the ERBB2 (0.066) and RET (0.054) mutations was remarkably improved for CRC.

**5.4. Mutation prediction comparison between the best-cluster optimized model and two baseline algorithms**

Figure 12 shows that our proposed best-cluster method outperformed both baseline algorithms proposed by Dooley et al. (2018) and Zhu et al. (2017). The superiority over the Zhu et al. model suggests that the combining of clusters with an AUC higher than 0.5 reduces the predictive ability of the model. The cluster distribution algorithm of Dooley et al. achieved an AUC mostly around or lower than 0.5, indicating that the model did not perform well on the gene mutation data.

These results suggest that unsupervised clustering can facilitate the identification of patches with better predictive values and exclude patches that lack predictive information. Furthermore, as expected, the introduction of less predictive clusters reduced the performance of the model. As a result, our proposed best-cluster approach outperformed Zhu's method when the top clusters with an AUC greater than 0.5 were combined and used to construct the prediction model.

**5.5. Mutation prediction comparison between the best-cluster optimized model and the tumor area model**

Figure 13 shows the average AUC of the model trained on the best clusters and the model trained only on tumor patches. The best tissue cluster model generally outperformed models trained only on tumor regions. The predictions of RET and BRAF mutations were improved by 0.062 and 0.042, respectively. In addition, Table 4 shows that patches in cluster 1 (tumor + mucin) and cluster 4 (a mix of all tissue types) had a higher predictive performance when compared with models based on the tumor patches identified by the automatic CRC tissue classifier.

**6. Discussion**

WSIs are widely used in digital pathology to predict gene mutations, molecular subtypes, and clinical outcomes. Since WSIs are too large (Giga pixels) to fit on a GPU at once, they are usually split into small image patches for training neural networks and prediction models. However, since patch-level labels are usually not available, we cannot directly perform classification on each patch. Therefore, multiple instance learning is often implemented to develop prediction models for patients. It is commonly assumed that tumor regions carry the most predictive information. Therefore, the development of deep learning models for the prediction of genetic mutations on WSIs are usually based solely on tumor tiles. In this paper, we proposed an unsupervised clustering method to segment WSIs according to the different morphologic features. Additionally, we also aimed to identify the best tissue tiles for the training of deep learning models for the prediction of gene mutations in three different types of cancers.

We demonstrated that the different clusters possessed had different predictive abilities. In addition, the clustering of image patches facilitated the identification of predictive patches and therefore improved the prediction of gene mutations for all three cancer types (LUAD, HNSCC, and CRC from TCGA) when compared with a model trained on all patches obtained from WSIs. These results suggest that unsupervised clustering can facilitate the identification of patches with better predictive values and exclude patches that lacked predictive information. Furthermore, our proposed algorithm outperformed two recently published baseline algorithms based on leveraging unsupervised clustering. Finally, the unsupervised clustering-based deep learning mutation prediction models made use of resolved probability scoring to facilitate the identification of spatial locations from each cluster that are most likely to be related to specific genetic mutations. This method further highlighted the importance of evaluating the heterogeneity of the tumor microenvironment to predict gene mutations.

Image tiles from tumor regions of a WSI are usually selected for constructing deep-learning digital pathology models based on the assumption that tumor cells possess most of the predictive information. Our findings have shown that while this hypothesis applied for HNSCC in cluster 4, for the LUAD cohort, tumor-like image tiles seem to be less predictive of the ALK, KRAS, and KEAP1 mutations (Table 3). Similarly, for the CRC cohort, neither the tumor tiles (cluster 2) identified by the unsupervised clustering nor the tumor patches identified by the supervised classifier (Table 4) provided a superior prediction performance for the gene mutation status. Conversely, the tumor and mucin tiles (cluster 1) in CRC as well tiles with a mixed variety of tissue types (cluster 4) had a higher predictive performance for gene mutations when compared to tumor tiles (Table 4). This suggests that the selection of tumor regions on WSIs is not always the best way to identify patches for the prediction of gene mutations, and other tissue types in the tumor micro-environment may provide a better prediction ability for certain phenotypes than tumor tissues. Previous studies have also shown that the mucin-to-tumor area ratio is highly correlated with the consensus molecular subtypes, MSI status, and the expression of mucin-producing genes (Nguyen et al., 2022).

Finally, we also demonstrated that unsupervised clustering could help reduce the workload for pathologist-based manual annotation. We assumed that a limited number of tissue types are present in WSIs, and the repeated clustering of the tiles could separate individual tissue types based on their morphologic appearance. Additionally, through further clustering of each cluster, we selected a small number of tiles near the center of each subcluster (e.g., 4 tiles). Therefore the pathologist only had to annotate the selected clusters. We showed that this semi-automatic annotation approach could identify similar tissue types on CRC WSIs to those identified by an automatic tissue classifier for CRC. This technique could be used to improve the interpretability of the unsupervised clustering-based deep learning model.

**Acknowledgements**

The research of Zihan Chen, Xingyu Li, and Hong Zhang was partially supported by National Natural Science Foundation of China (No. 11771096, 72091212), Anhui Center for Applied Mathematics, and Special Project of Strategic Leading Science and Technology of CAS (No. XDC08010100).

**Author contributions**

X.S.X., X.L., Z.C., and H.Z. contributed to design of the research; X.L., Z.C., and X.S.X. contributed to data acquisition; X.L., Z.C., and X.S.X. contributed to data analysis. Z.C., X.L., X.S.X., M.Y., and H.Z. contributed to data interpretation. Z.C., X.S.X., X.L., and H.Z. wrote the manuscript; and all authors critically reviewed the manuscript and approved the final version

**Data availability**

The TCGA dataset is publicly available at the TCGA portal (https://portal.gdc.cancer.gov).The public TCGA clinical data is available at the website(https://xenabrowser.net/datapages/). Xception model weights are available at (https://github.com/fchollet/deep-learningmodels /releases/download/v0.4/xception_weights_tf_dim_ordering_tf_kernels_notop.h5).

**Code availability**

Source code is available at https://github.com/ChenZHUSTC/BCOMIL.

**Tables**

**Table 1.** Numbers of mutation and wild type of each gene in three cancers.

| Gene | Total | Mutant | Wild Type | Mutation Frequency |
|---|---|---|---|---|
| LUAD | | | | |
| TP53 | 434 | 224 | 210 | 0.52 |
| STK11 | 434 | 63 | 371 | 0.15 |
| KEAP1 | 434 | 77 | 357 | 0.18 |
| EGFR | 434 | 55 | 379 | 0.13 |
| ALK | 434 | 24 | 410 | 0.06 |
| KRAS | 434 | 132 | 302 | 0.30 |
| HNSCC | | | | |
| TP53 | 431 | 320 | 111 | 0.74 |
| CASP8 | 431 | 47 | 384 | 0.11 |
| NSD1 | 431 | 51 | 380 | 0.12 |
| HRAS | 431 | 27 | 404 | 0.06 |
| PTEN | 431 | 10 | 421 | 0.02 |
| DNAH5 | 431 | 58 | 373 | 0.13 |
| CRC | | | | |
| TP53 | 414 | 260 | 154 | 0.63 |
| PIK3CA | 414 | 158 | 256 | 0.38 |
| ATM | 414 | 120 | 294 | 0.29 |
| MET | 414 | 38 | 376 | 0.09 |
| BRAF | 414 | 91 | 323 | 0.22 |
| RET | 414 | 27 | 387 | 0.07 |
| ERBB2 | 414 | 30 | 384 | 0.63 |

**Table 2.** Average AUC (standard deviation) from 5-fold cross validation for different clusters in TCGA LUAD

| Gene Cluster | TCGA (5-fold CV) | | | |
|---|---|---|---|---|
| | Cluster 1 (N =433) | **Cluster 2 (N = 428)** | Cluster 3 (N = 434) | Cluster 4 (N = 434) |
| **TP53** | 0.655±0.077 | **0.692±0.082** | 0.609±0.070 | 0.584±0.075 |
| **STK11** | 0.608±0.095 | **0.647±0.100** | 0.553±0.100 | 0.563±0.157 |
| **EGFR** | **0.649±0.126** | 0.643±0.118 | 0.584±0.107 | 0.595±0.123 |
| **ALK** | 0.549±0.192 | 0.609±0.151 | 0.544±0.123 | **0.655±0.233** |
| **KRAS** | 0.517±0.053 | 0.536±0.054 | **0.608±0.068** | 0.564±0.070 |
| **KEAP1** | **0.630±0.137** | 0.594±0.152 | 0.619±0.084 | 0.611±0.081 |

**Table 3.** Average AUC (standard deviation) from 5-fold cross validation for different clusters in TCGA HNSCC

| Gene | TCGA (5-fold CV) | | | |
|---|---|---|---|---|
| **Cluster** | Cluster 1 (N =431) | Cluster 2 (N = 431) | Cluster 3 (N = 430) | **Cluster 4 (N = 430)** |
| **TP53** | 0.690±0.073 | 0.611±0.128 | 0.596±0.068 | **0.719±0.061** |
| **DNAH5** | 0.462±0.090 | **0.604±0.064** | 0.505±0.067 | 0.479±0.088 |
| **HRAS** | 0.590±0.152 | **0.665±0.140** | 0.454±0.103 | **0.658±0.178** |
| **CASP8** | **0.666±0.124** | 0.564±0.105 | 0.638±0.061 | **0.665±0.072** |
| **PTEN** | 0.540±0.286 | **0.625±0.230** | 0.577±0.204 | 0.552±0.225 |
| **NSD1** | 0.630±0.100 | 0.632±0.121 | **0.657±0.089** | **0.639±0.070** |

**Table 4.** Average AUC (standard deviation) from 5-fold cross validation for different clusters in TCGA CRC

| Gene | TCGA (5-fold CV) | | | | |
|---|---|---|---|---|---|
| **Cluster** | **Cluster 1** (N =413) | Cluster 2 (N = 411) | Cluster 3 (N = 414) | Cluster 4 (N = 414) | Tumor patches* |
| **TP53** | **0.657±0.034** | 0.642±0.059 | 0.575±0.061 | **0.653±0.028** | 0.665±0.043 |
| **PIK3CA** | **0.759±0.071** | 0.706±0.083 | 0.721±0.081 | **0.766±0.048** | 0.737±0.085 |
| **BRAF** | **0.729±0.043** | 0.666±0.066 | 0.625±0.072 | **0.703±0.078** | 0.687±0.054 |
| **ERBB2** | 0.554±0.145 | 0.551±0.135 | 0.546±0.049 | **0.633±0.167** | 0.598±0.137 |
| **ATM** | **0.738±0.036** | 0.733±0.056 | 0.720±0.030 | **0.743±0.026** | 0.734±0.054 |
| **MET** | **0.703±0.097** | 0.695±0.062 | 0.696±0.096 | **0.737±0.112** | 0.697±0.052 |
| **RET** | **0.685±0.094** | 0.529±0.111 | 0.510±0.123 | **0.677±0.076** | 0.623± 0.048 |

Note: * model trained on tumor patches identified by a tissue classifier for CRC.

**Figure 1. Framework of unsupervised clustering-based deep-learning modeling for prediction of gene mutations.** Images of three cancer tissues were first downloaded from TCGA and CPTAC databases. Each whole-slide H&E image was preprocessed to (1) remove the background areas using a U-net, (2) split into non-overlapping tiles with a size of 224 x 224 pixels, and (3) color normalized. A fine-tuned Xception model-based feature extractor was used to generate patch representations. For each cancer type, K-means clustering was used to group patches into four clusters. Neural networks with the same structure were trained on each cluster data and all-patch data to obtain patch-level classifiers. The results were finally aggregated per-slide to produce the heatmaps and the AUC statistics.

**Figure 2. Framework of semi-automatic annotation of clusters.** An additional K-means clustering (k = 4) was performed, and four neighborhood patches around the center of each subcluster were selected for pathologists' annotation.

**Figure 3. t-SNE visualization of clustering results.** For each cancer, 5,000 patches were randomly selected from each of the four clusters and displayed using t-distributed stochastic neighbor embedding (t-SNE) dimensionality reduction representation.

**Figures 4. Representative Image tiles for each cluster for LUAD.** For each subcluster, a representative patch is displayed. The tissue type for each tile is annotated using a semi-automatic annotation approach.

**Figures 5. Representative Image tiles for each cluster for HNSCC.** For each subcluster, a representative patch is displayed. The tissue type for each tile is annotated using a semi-automatic annotation approach.

**Figures 6. Representative Image tiles for each cluster for LUAD.** For each subcluster, a representative patch is displayed. The tissue type for each tile is annotated using a semi-automatic annotation approach (a) and a CRC tissue-type classifier (b).

**Figure 7. Sankey diagram of CRC clustering results.** Sankey diagram is used to identify tissue types in different clusters by comparing with the tissue types predicted from a CRC tissue classifier.

**Figures 8. Visualization of the proposed algorithm for different genes in LUAD.** The deep learning-based unsupervised clustering and mutation predictions are visualized to understand the spatial locations of each cluster, to identify the spatial regions related to mutation of a specific gene via the resolved probability scores, and to highlight the heterogeneity of a predicted genotype in the tumor microenvironment. The heatmap showcases the probability scores of the gene mutations in the identified best cluster. The tile with the highest probability of mutations for each gene is displayed and the corresponding tissue type is provided.

**Figures 9. Visualization of the proposed algorithm for different genes in HNSCC.** The deep learning-based unsupervised clustering and mutation predictions are visualized to understand the spatial locations of each cluster, to identify the spatial regions related to mutation of a specific gene via the resolved probability scores, and to highlight the heterogeneity of a predicted genotype in the tumor microenvironment. The heatmap showcases the probability scores of the gene mutation in the identified best cluster. The tile with the highest probability of mutation for each gene is displayed and the corresponding tissue type is provided.

**Figures 10. Visualization of the proposed algorithm for different genes in CRC.** The deep learning-based unsupervised clustering and mutation predictions are visualized to understand the spatial locations of each cluster, to identify the spatial regions related to mutation of a specific gene via the resolved probability scores, and to highlight the heterogeneity of a predicted genotype in the tumor microenvironment. The heatmap showcases the probability scores of the gene mutation in the identified best cluster. The tile with the highest probability of mutation for each gene is displayed and the corresponding tissue type is provided.

**Figure 11. Comparison of model performance (average AUC score) of the proposed clustering-based algorithm with models using whole-slide images without patch selection.** Red points = the best cluster results; green points = models using whole-slide images. The bar charts showcase the difference in average AUC between clustering model and all-patch model.

**Figure 12. Comparison of model performance (average AUC score) of the proposed clustering-based algorithm with two baseline methods (Dooley et al.'s method and Zhu et al.'s method)**

**Figure 13. Comparison of the proposed BCOMIL method with model trained on tumor patches in CRC.** Red points = the best cluster results; green points = model trained on tumor patches. The bar charts showcase the difference in average AUC between clustering model and all-patch model.

Figure 1

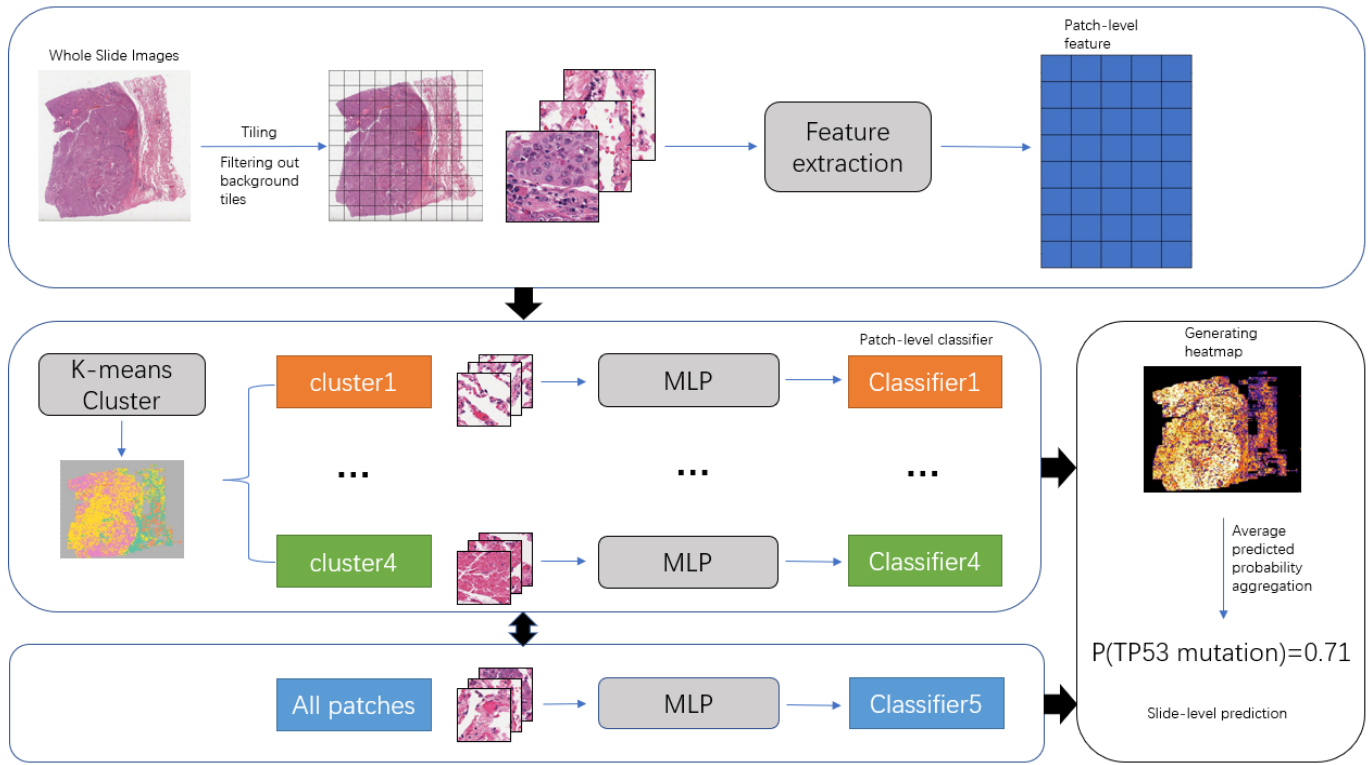

Figure 2.

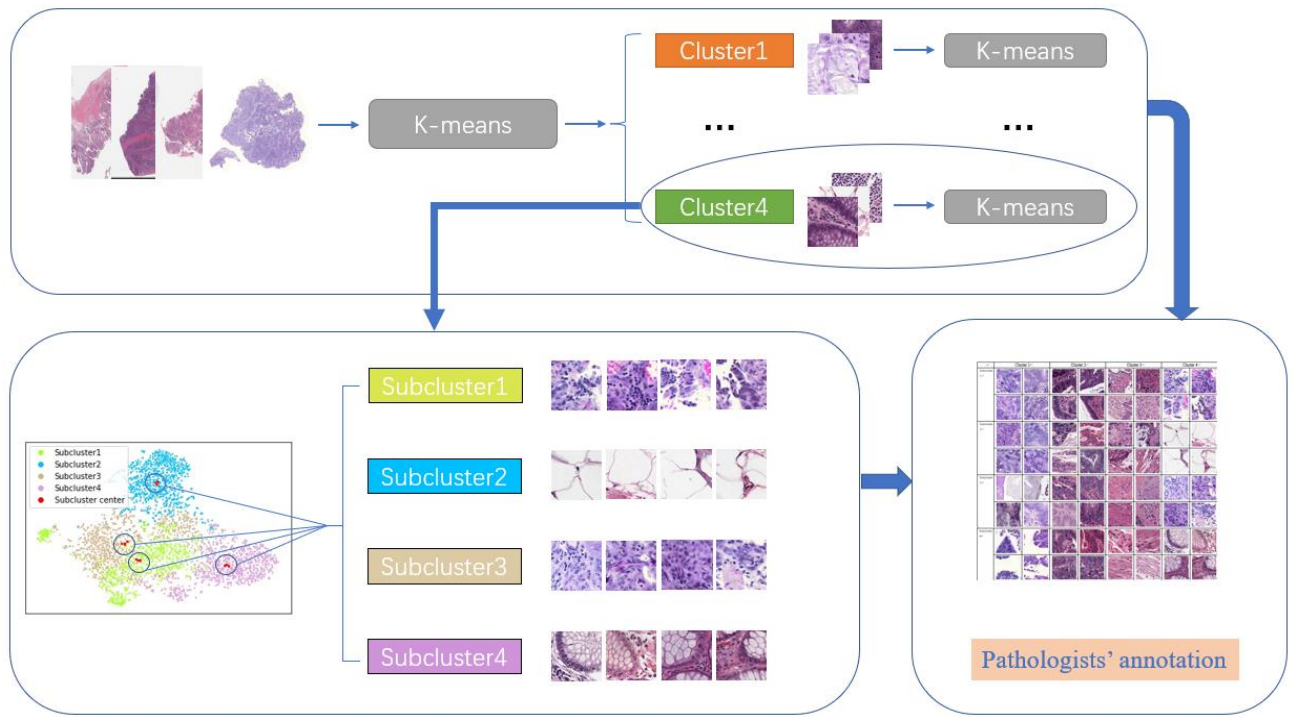

Pathologists' annotation

Figure 3.

| LUAD | HNSCC | CRC |
|---|---|---|
| 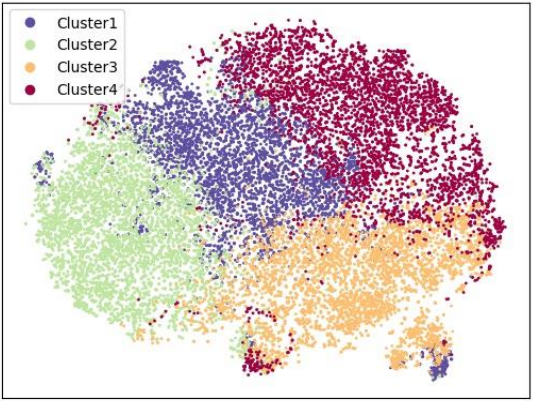 | 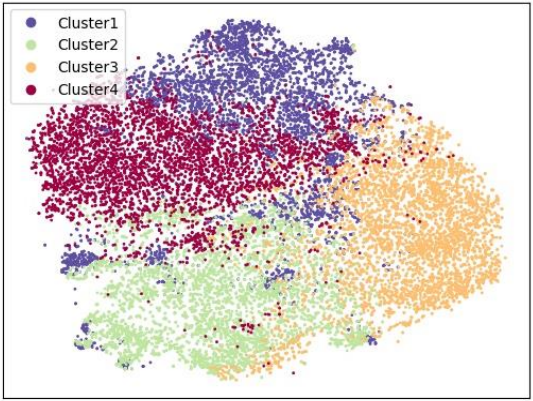 | 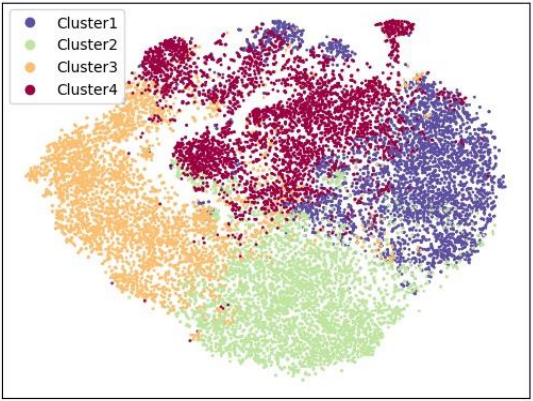 |

Figure 4.

| | Cluster 1 | Cluster 2 | Cluster 3 | Cluster 4 |
|---|---|---|---|---|
| Subcluster 1 | Stroma, Atypical cells 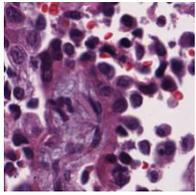 | Tumor 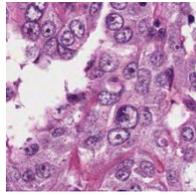 | Stroma, Atypical cells 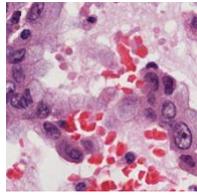 | Stroma 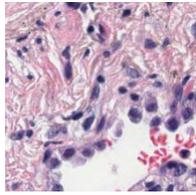 |
| Subcluster 2 | Pulmonary alveolus 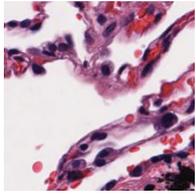 | Tumor 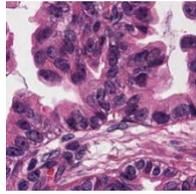 | Pulmonary alveolus 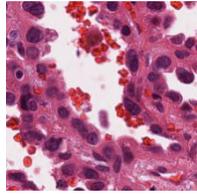 | Stroma 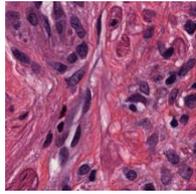 |
| Subcluster 3 | Suspected tumor cells 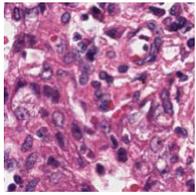 | Tumor 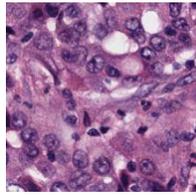 | Lymphocytes/ proliferating fibroblasts 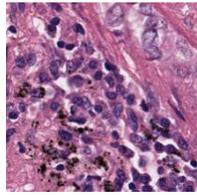 | Stroma 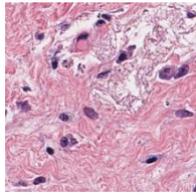 |
| Subcluster 4 | Tumor 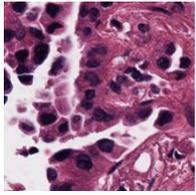 | Tumor 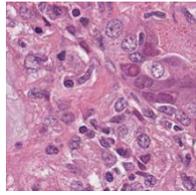 | Non-tumor 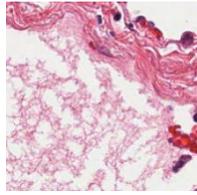 | Stroma/ Lymphocytes 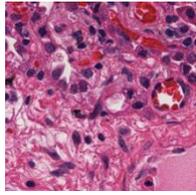 |

Figure 5.

| | Cluster 1 | Cluster 2 | Cluster 3 | Cluster 4 |
|---|---|---|---|---|
| Subcluster 1 | Tumor 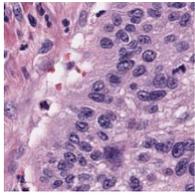 | Non-tumor Blood vessels 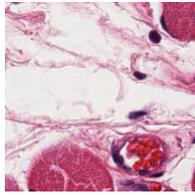 | Non-tumor 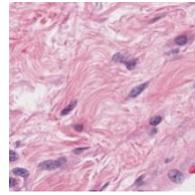 | Tumor 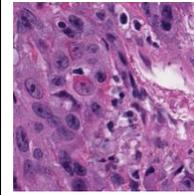 |
| Subcluster 2 | Lymphocytes 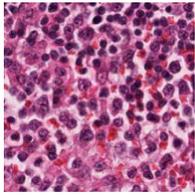 | Non-tumor 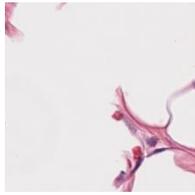 | Non-tumor 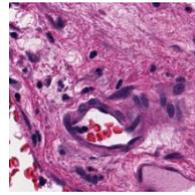 | Tumor 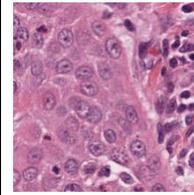 |
| Subcluster 3 | Tumor 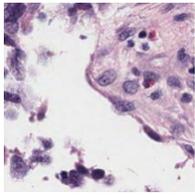 | Tumor 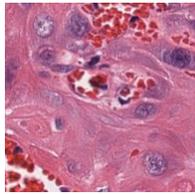 | Non-tumor 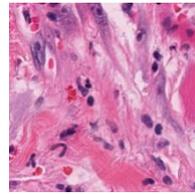 | Tumor 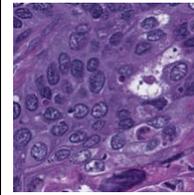 |
| Subcluster 4 | Tumor 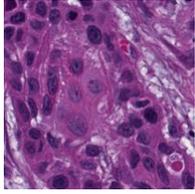 | Necrotic debris 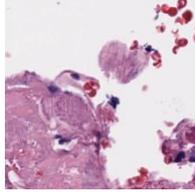 | Muscle, adipose 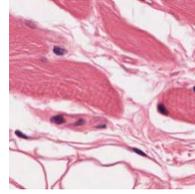 | Tumor 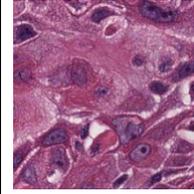 |

Figure 6.(a)

|  | Cluster 1 | Cluster 2 | Cluster 3 | Cluster 4 |
|---|---|---|---|---|
| Subcluster 1 | Tumor 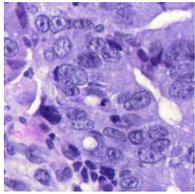 | Tumor 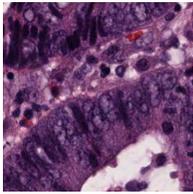 | Debris 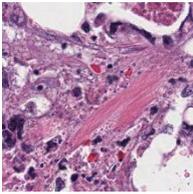 | Lymphocytes 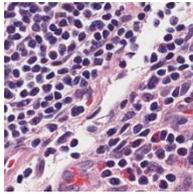 |
| Subcluster 2 | Tumor/Mucin 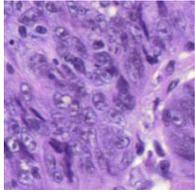 | Tumor 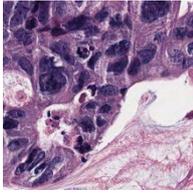 | Non-tumor 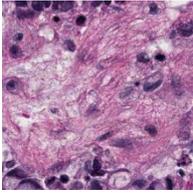 | Adipose 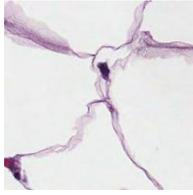 |
| Subcluster 3 | Mucin 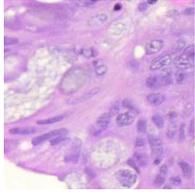 | Tumor 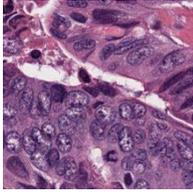 | Stroma 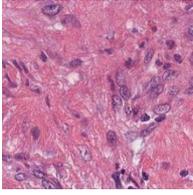 | Tumor 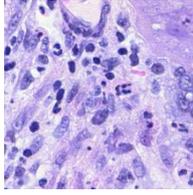 |
| Subcluster 4 | Tumor 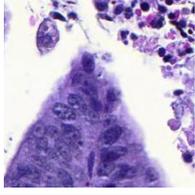 | Tumor 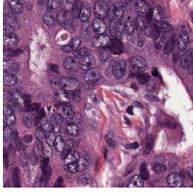 | Muscule 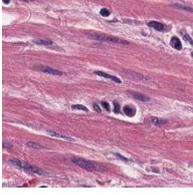 | Non-tumor 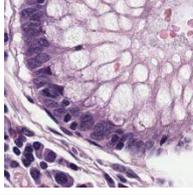 |

(b)

| | Cluster 1 | Cluster 2 | Cluster 3 | Cluster 4 |
|---|---|---|---|---|
| Subcluster 1 | Tumor 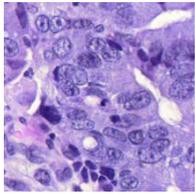 | Tumor 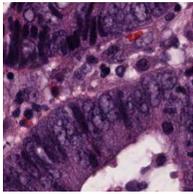 | Debris 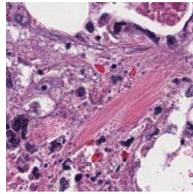 | Lymphocytes 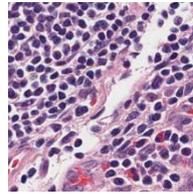 |
| Subcluster 2 | Mucin 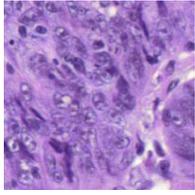 | Tumor 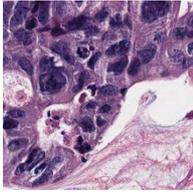 | Stroma 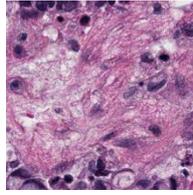 | Adipose 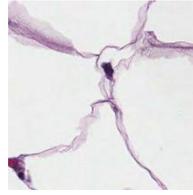 |
| Subcluster 3 | Mucin 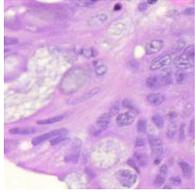 | Tumor 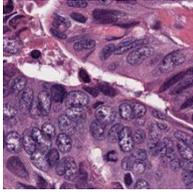 | Stroma 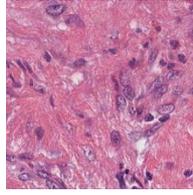 | Mucin 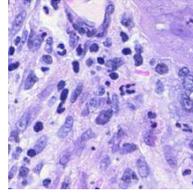 |
| Subcluster 4 | Tumor 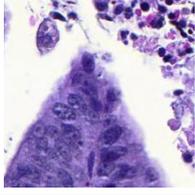 | Tumor 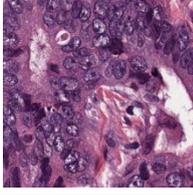 | Muscle 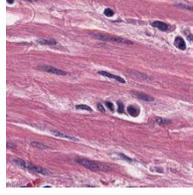 | Normal 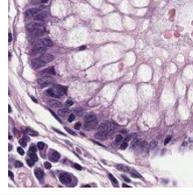 |

Figure 7.

Figure 8.

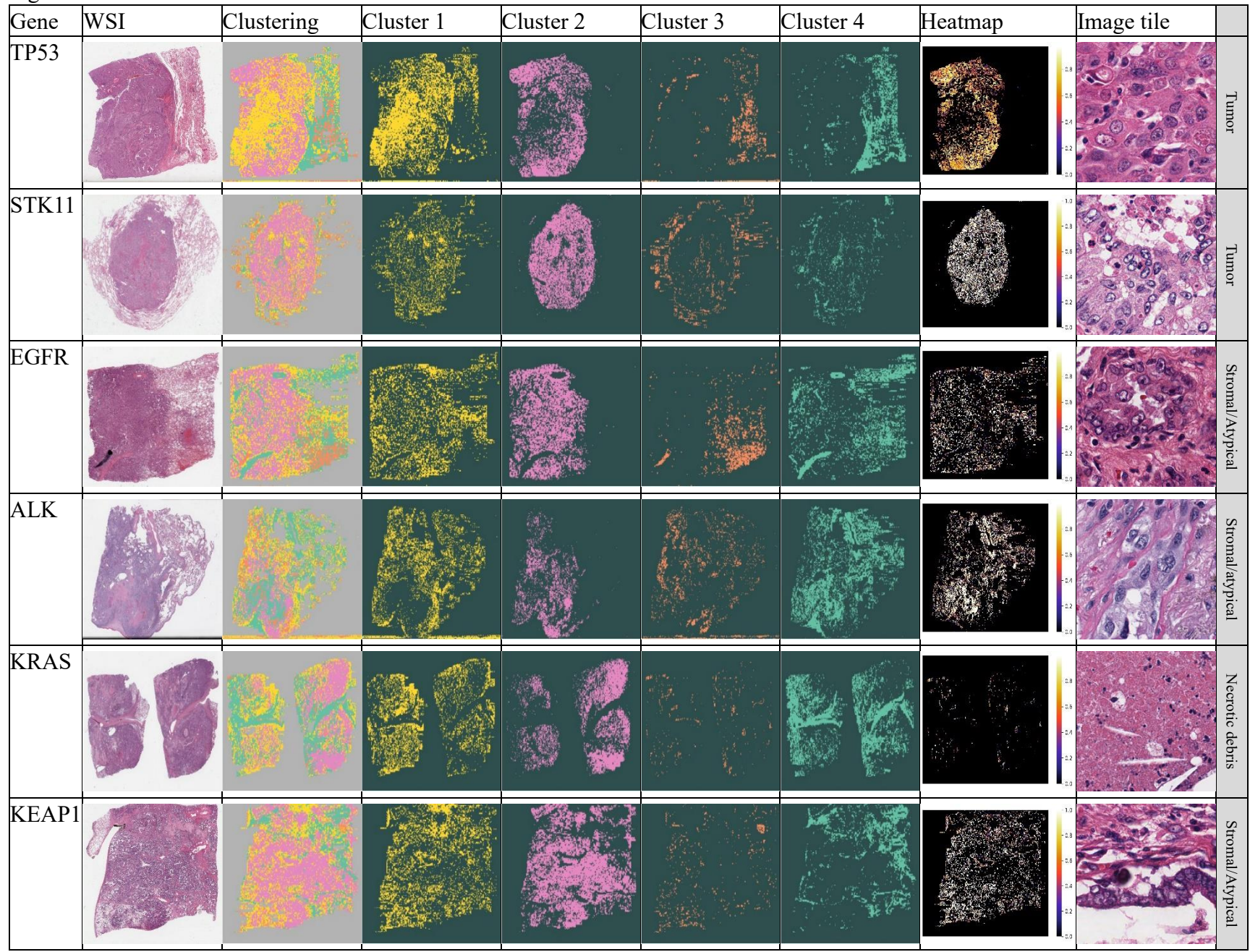

Figure 9.

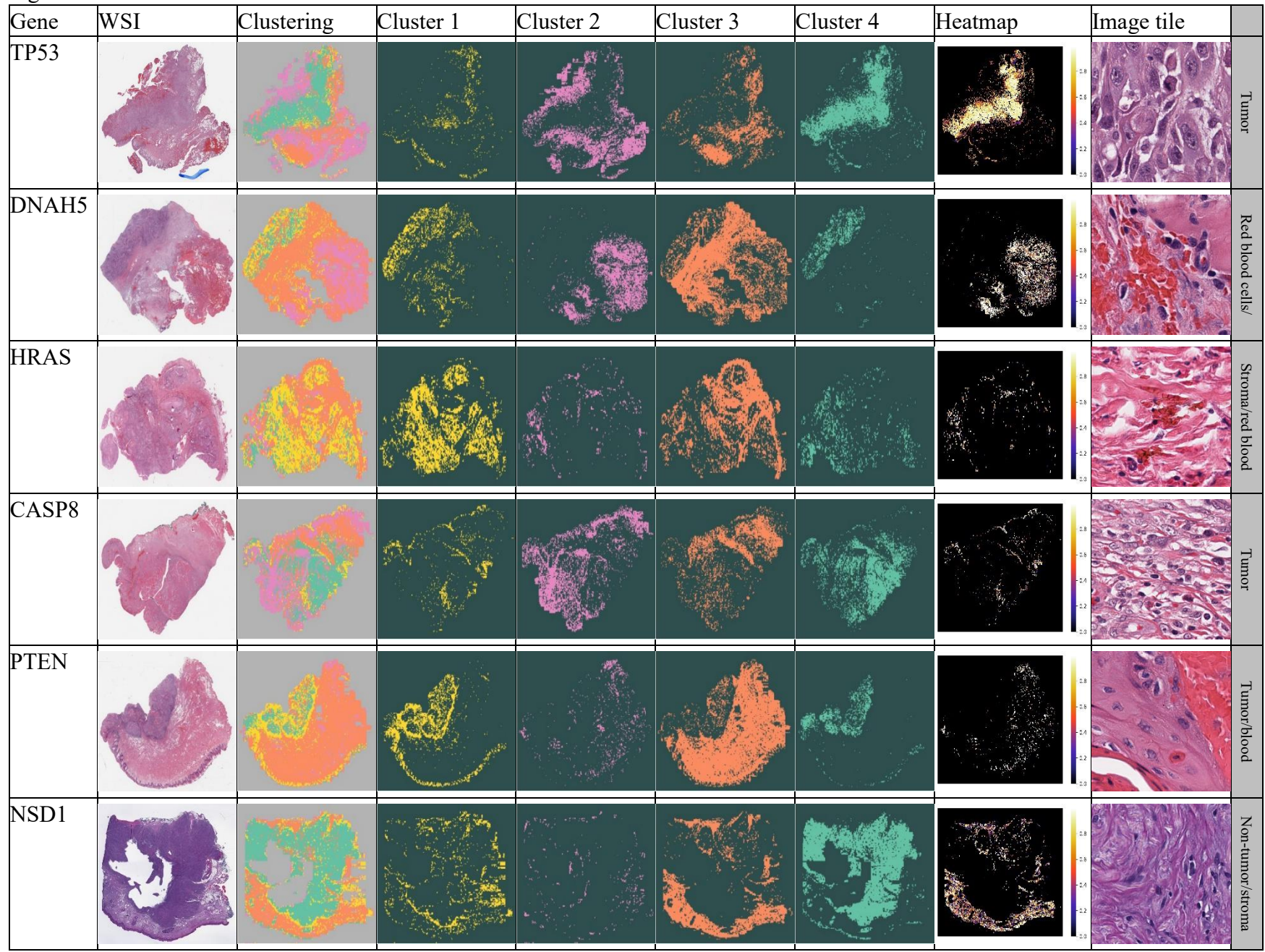

Figure 10.

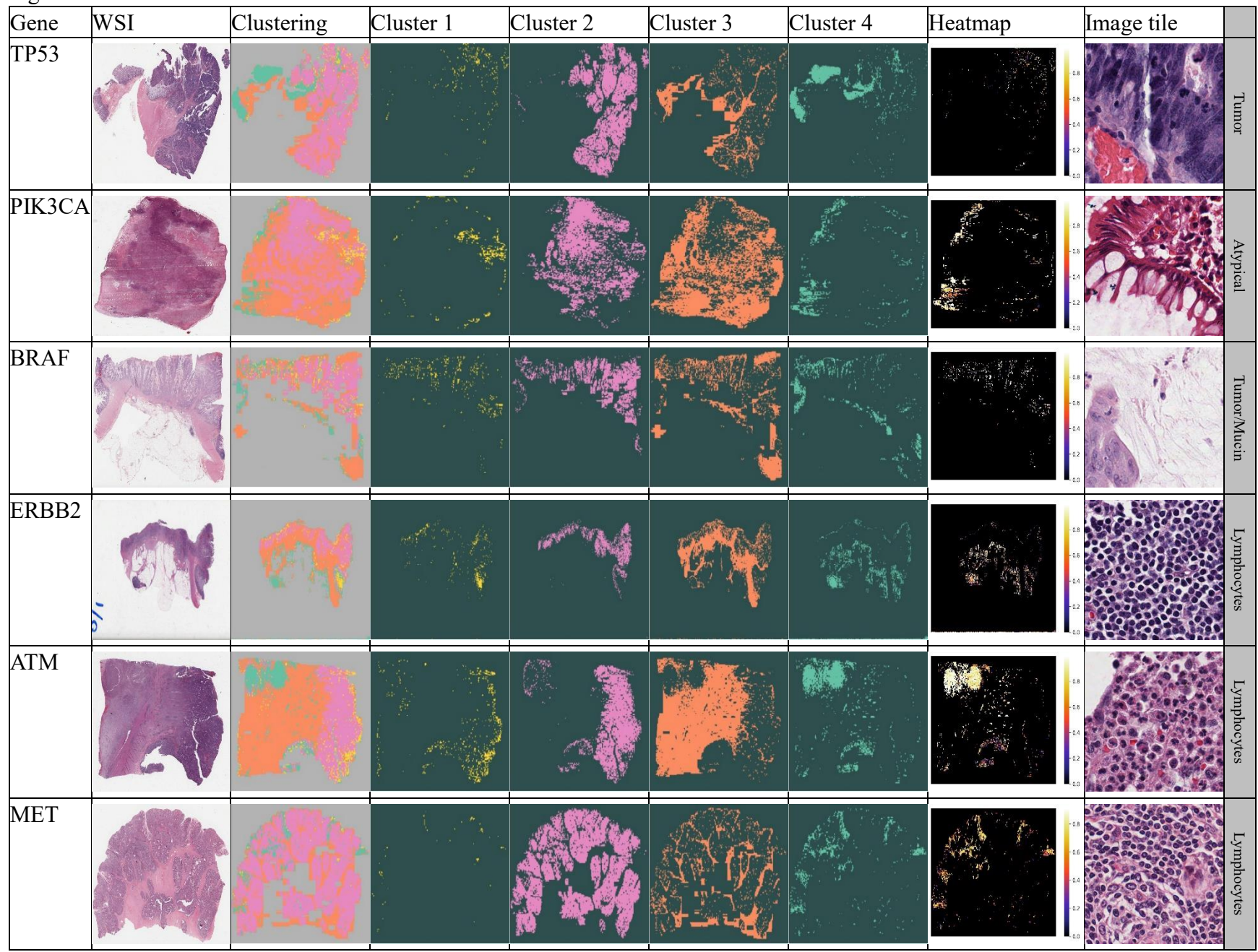

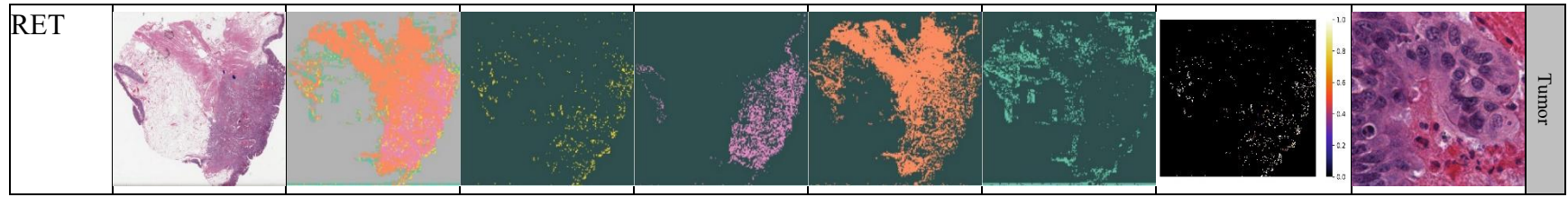

Figure 11.

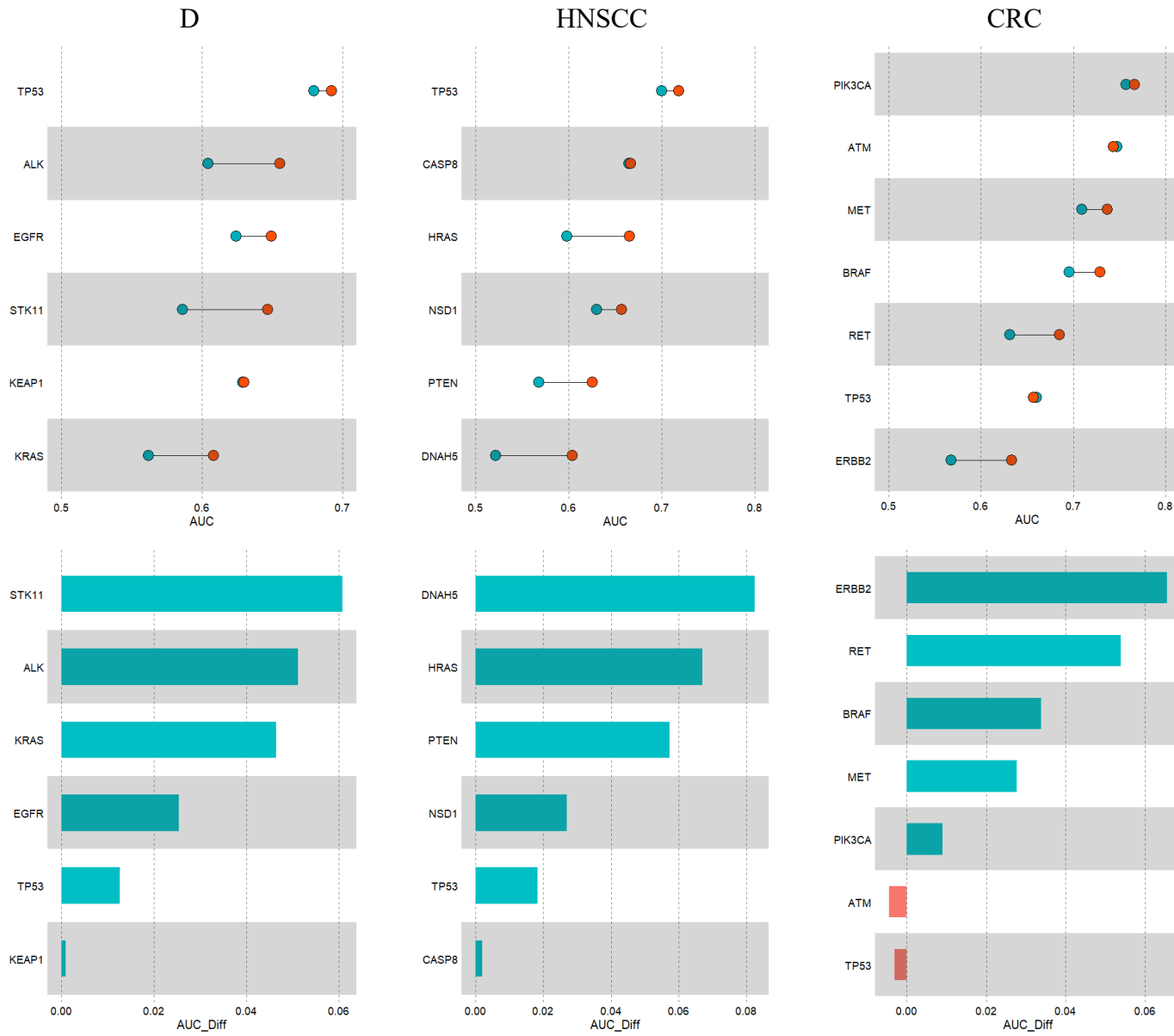

Figure 12.

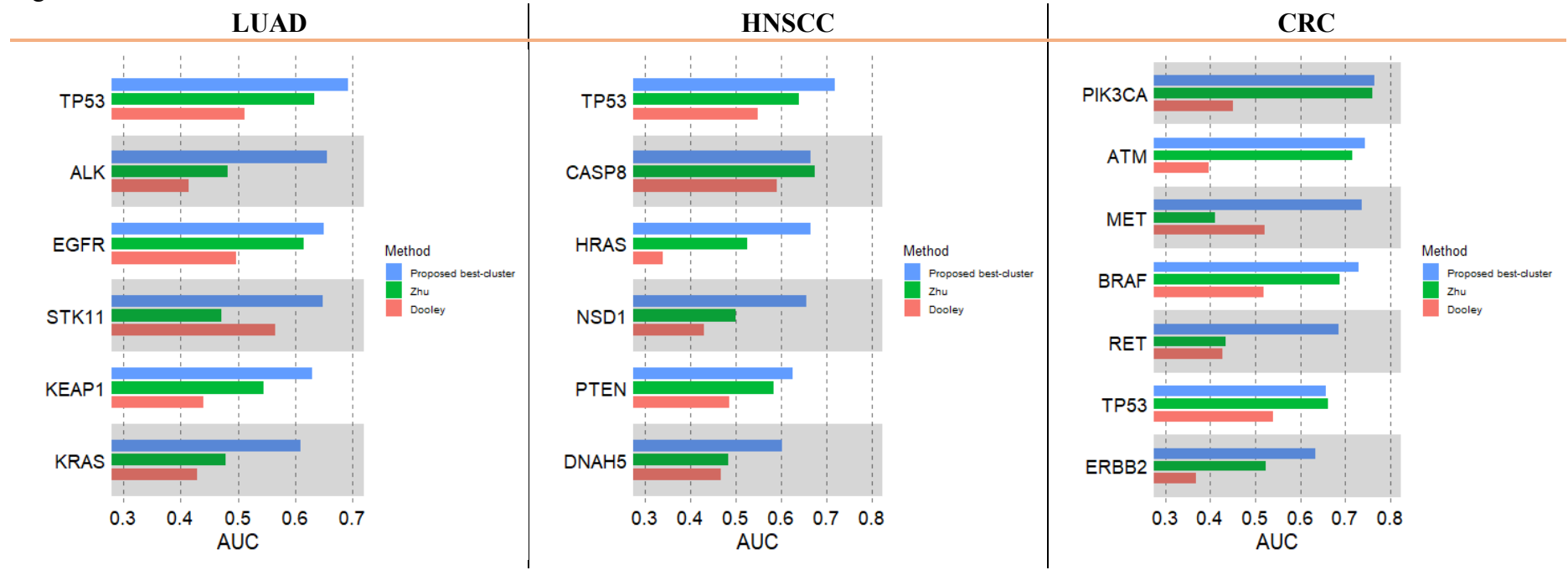

Figure 13.

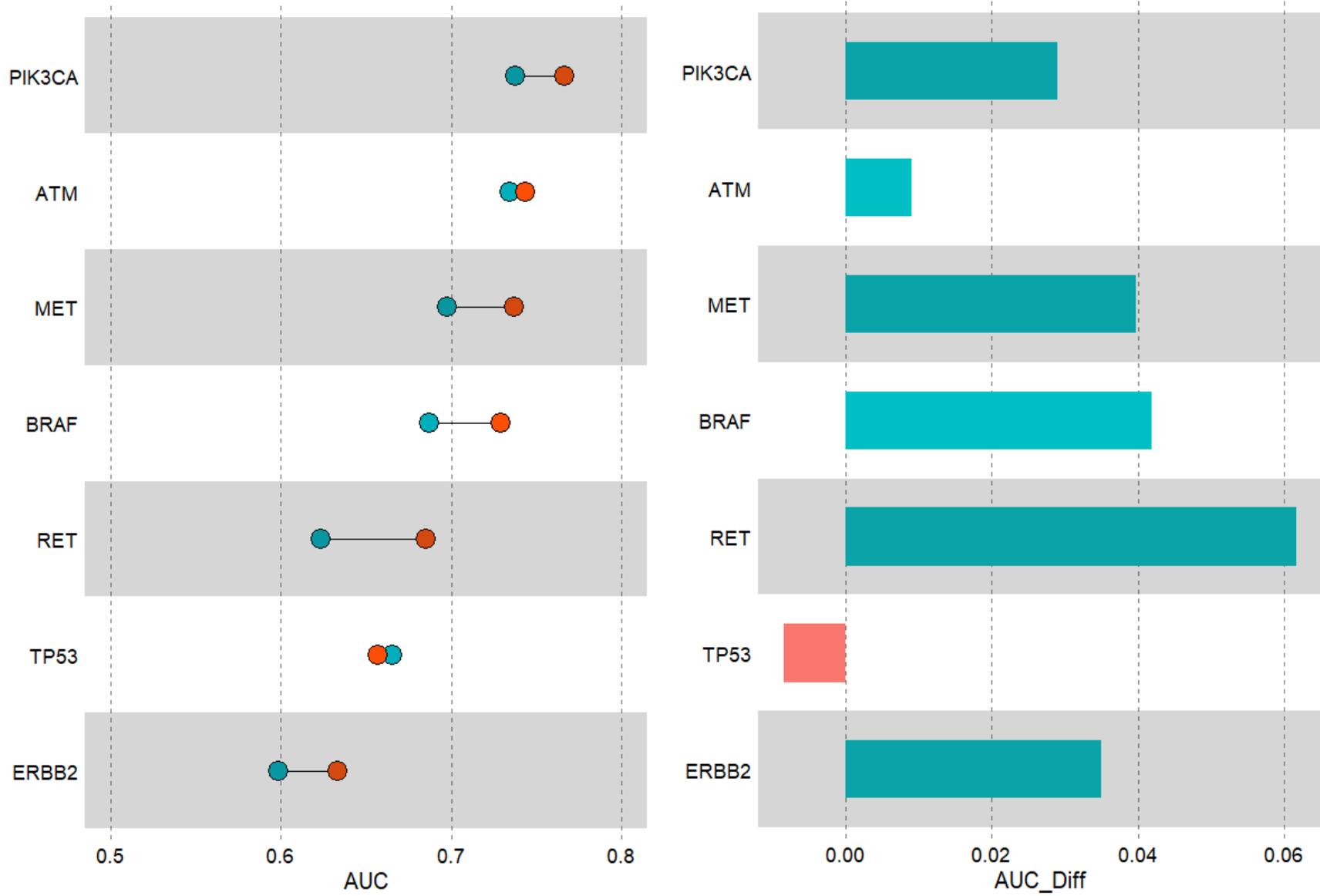

**Supplementary:**

Supplementary Table 1. A summary of the number of patients of each cancer in different tasks.

| cancer | LUAD mutation | HNSCC mutation | CRC mutation |
|---|---|---|---|
| tcga_all | 434 | 431 | 414 |
| tcga_c1 | 433 | 431 | 413 |
| tcga_c2 | 428 | 431 | 411 |
| tcga_c3 | 434 | 430 | 414 |
| tcga_c4 | 434 | 430 | 414 |

Supplementary Figure 1. Target patch of Macenko's normalization

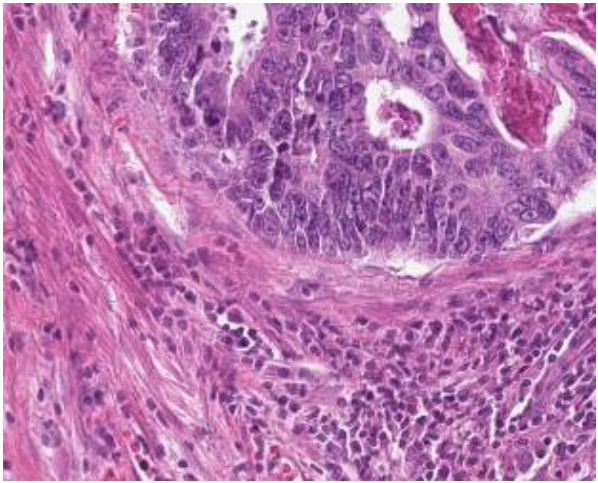

Supplementary Figure 2. Confusion matrix of CRC tissue-type classification